\documentclass{webofc}

\usepackage[varg]{txfonts}   
\usepackage{hyperref}
\usepackage{url}
\hypersetup{colorlinks=true,citecolor=blue,urlcolor=blue,linkcolor=blue}
\begin{document}
\title{Molecular states with charm: insights from vacuum and finite-temperature analyses}
%
%

\author{\firstname{Juan M.} \lastname{Torres-Rincon}\inst{1}\fnsep\thanks{\email{torres@fqa.ub.edu} 
}}

\institute{Departament de F\'isica Qu\`antica i Astrof\'isica and Institut de Ci\`encies del Cosmos (ICCUB), Facultat de F\'isica,  Universitat de Barcelona, Mart\'i i Franqu\`es 1, 08028 Barcelona, Spain}

\abstract{This contribution to the SQM2024 conference covers the molecular hypothesis for the internal structure of some open and hidden charm states. The use of effective theories that incorporate heavy-quark spin symmetry, combined with unitarization techniques, has provided strong evidences  supporting this interpretation. In the heavy-light sector, we discuss the double pole structure of the $D_0^*(2300)$ and the generation of the $D_{s0}^*(2317)$. In the hidden charm sector, we focus on the exotic $X(3872)$ and its heavy-quark partner, the $X(4014)$. Furthermore, we emphasize the benefits of femtoscopic measurements in $p+p$ collisions to establish the nature of these states, as well as the potential role of temperature to discern their internal structure.
}
\maketitle
\section{Introduction}
\label{intro}

Since the advent of the quark model, observed hadrons were usually classified as composite objects consisting of valence quarks and antiquarks. Following  the developments of QCD, the inclusion of the gluon as a possible valence component expanded the possibilities for combining these pieces to form color neutral particles. In the literature various composite structures can be classified as (for recent reviews, see~\cite{Chen:2016spr,Guo:2017jvc,Brambilla:2019esw,Chen:2022asf,Liu:2024uxn}),

\begin{itemize}
    \item Conventional hadrons: mesons ($\bar{q}q$) and baryons ($qqq$).
    \item Glueballs: $gg$, $ggg$...
    \item Multiquark states: tetraquarks, pentaquarks...
    \item Hybrids: $qqqg$, $q\bar{q}g$...
    \item $\cdots$
\end{itemize}

In this contribution, we will focus on another interpretation known as {\it hadronic molecules}, in which colorless states---such as standard mesons and baryons---are bound by the residual strong force. Their interactions are typically modeled using an appropriate effective field theory (EFT), and a unitarization (or a two-body equation) technique.

In addition, we will consider states that contains at least one charm quark, for which heavy-quark effective theory can serve as an initial starting point. We will explore open-charm as well as hidden-charm (exotic) states, all of them within the molecular hypothesis. Predictions for these states from EFTs exist not only in vacuum, but also at finite temperatures, where the properties of the molecular states, such as masses and decay widths, are modified by the medium. In fact, even the nature of the states at $T=0$ can change when immersed in the thermal bath. This is particularly relevant in the context of relativistic heavy-ion collisions (RHICs) where hadron production starts at around $T \simeq 160$ MeV, until they freeze out at a lower temperature. Therefore the properties of these states and the subsequent decay products can be affected by the temperature.

\section{Heavy-light states}
\label{sec:heavylight}

Consider the lightest charm meson, $D$, and its heavy-quark partner, $D^*$, forming composite states with light mesons through their attractive interactions. These can be described by an EFT that incorporates chiral and heavy-quark spin symmetries~\cite{Kolomeitsev:2003ac,Hofmann:2003je,Guo:2006fu,Guo:2008gp,Tolos:2013kva,Altenbuchinger:2013vwa}. Other approaches extend the chiral symmetry to four flavors~\cite{Gamermann:2006nm}. For a recent review, see Ref.~\cite{Das:2024vac}.

\begin{figure*}
\centering
\vspace*{1cm}       
\includegraphics[width=4cm,clip]{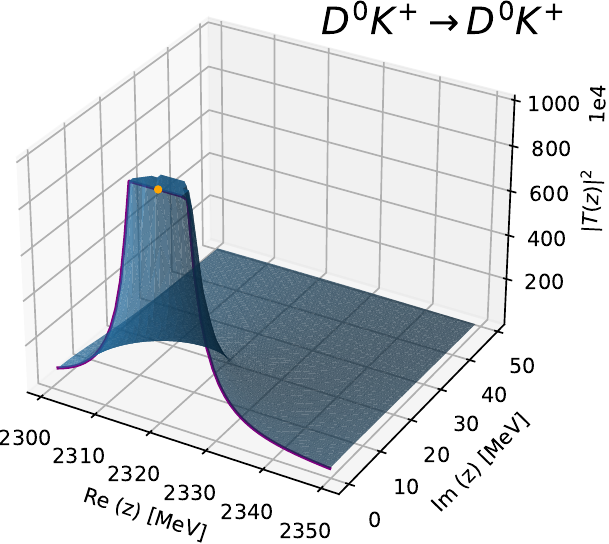}
\includegraphics[width=4cm,clip]{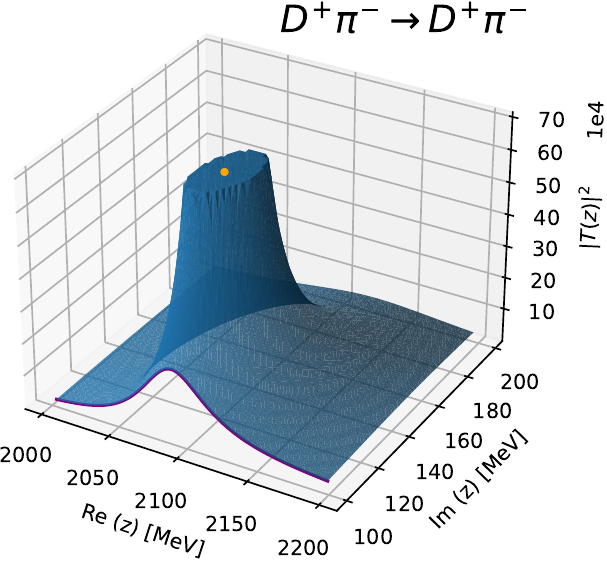}
\includegraphics[width=4cm,clip]{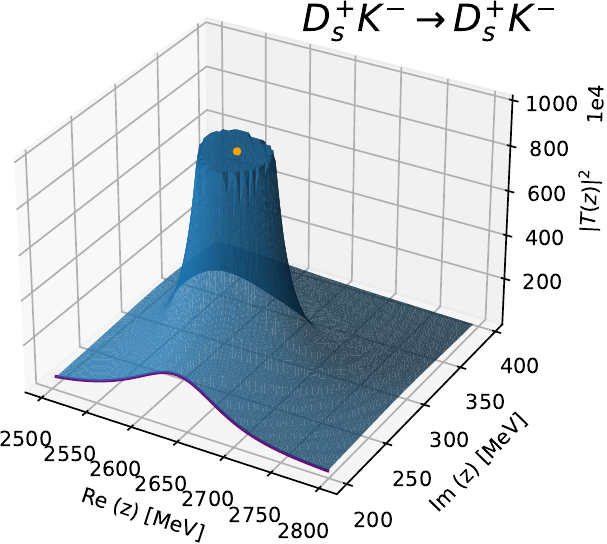}
\caption{Left panel: Pole of the $T$ matrix corresponding to the physical $D_{s0}^*(2317)$ bound state. Middle and right panels: two poles of the single $D_0^*(2300)$ molecular state, shown in different reactions of the same coupled-channels problem. Figures obtained from the results given in Ref.~\cite{Torres-Rincon:2023qll}.}
\label{fig:D02300}       
\end{figure*}

For example, the narrow $D_{s0}^* (2317)$ ~\cite{BaBar:2003oey} can be understood as a bound state of the $D^0K^+$ interaction (and its coupled channels) in the EFT approach~\cite{Kolomeitsev:2003ac,Hofmann:2003je,Guo:2006fu} (it contains a small decay width due to the isospin-breaking decay mechanism). In the left panel of Fig.~\ref{fig:D02300} we show the $T$-matrix pole associated to this state. The $D_{s1} (2460)$ corresponds to the heavy-quark spin partner, replacing the $D$ meson with the vector $D^*$. Other exotic states with similar masses have also been explained using meson-meson dynamics, such as the $X_0(2866)$, interpreted as a $D^* \bar{K}^{*-}$ molecule~\cite{Liu:2020nil}.

An interesting case arises with two rather broad states, the $D_0^*(2300)$ and $D_1^*(2430)$, which were difficult to reconcile with the results of EFTs given their experimental masses and widths. The solution of this puzzle came through the hypothesis of a ``double-pole structure'', where two different poles of the $T$ matrix emerge with the same quantum numbers and interacting channels~\cite{Albaladejo:2016lbb,Meissner:2020khl,Torres-Rincon:2021wyd}. Depending on their partial couplings, these poles present more strength in different final channels. In Fig.~\ref{fig:D02300} we show the $T$-matrix element of the $D^+ \pi^- \rightarrow D^+ \pi^-$ reaction (middle panel), where the lower pole is identified; and the $T$ matrix of the $D_s^+ K^- \rightarrow D_s^+ K^-$ reaction (right panel), where the higher pole emerges. In Ref.~\cite{Albaladejo:2016lbb} the two poles are separated by 250 MeV in mass, both with large decay widths (> 200 MeV).

Meson-baryon molecular candidates are also supported by EFT calculations. Examples where the $s-$wave interactions produce multiparticle states include the $\Lambda_c(2595)$, $\Lambda_c (2880)$, $\Sigma_c(2800)$ and $\Xi_c (2970)$ (among many others). See for example~\cite{Hofmann:2003je,Mizutani:2006vq,He:2010zq,He:2006is}, as well as \cite{Guo:2017jvc} for a more comprehensive review.

\section{Heavy-heavy states}
\label{sec:heavy-heavy}

Molecular states containing double charm have also been explored within the same framework. In this case the bound states can be generated by two $D/D^*$ mesons, or by a $D/D^*$ meson and a $\bar{D}/\bar{D}^*$ meson, in the hidden charm case. The exotic $X(3872)$, first measured at Belle~\cite{Belle:2003nnu}, has been proposed to admit different interpretations, including tetraquark, conventional $\chi_{c1}(2P)$, hybrid, and others~\cite{Liu:2024uxn}. The molecular possibility is particularly appealing since the mass of the $X(3872)$ is very close to the $D\bar{D}^*$ threshold, with an extremely small width. This argument is backed up by the explicit calculations using SU(4) chiral theory~\cite{Gamermann:2007fi} as well as the local hidden gauge formalism~\cite{Molina:2009ct}. Using this EFT, Ref.~\cite{Montana:2022inz} generated both the $X(3872)$ and its heavy-quark partner, the $X(4014)$. 

Many exotic states (traditionally denoted as $X,Y,Z$) have been discovered in the last years, with some of them lying close to open-flavor thresholds and postulated to be dominated by a molecular component. For further details on these states see Refs.~\cite{Guo:2017jvc,Liu:2024uxn}. Some insights can also be gained from the heavy-ion program. For example, the molecular and tetraquark configurations of the $X(3872)$ would yield very different production rates as a function of the collision centrality~\cite{Zhang:2020dwn}. The first measurement of this state in RHICs was reported in~\cite{CMS:2021znk}.

\section{Charm femtoscopy in $p$-$p$ collisions}

The femtoscopy technique~\cite{Heinz:1999rw,Lisa:2005dd,Fabbietti:2020bfg} has gained a lot of attention in recent years thanks to the experimental capabilities for measuring correlation functions in high-energy $p$+$p$ collisions and the variety of hadrons that can be studied. Being sensitive to the final interactions of the measured pair, femtoscopy is able to help in identifying possible bound states or resonances formed by the detected hadron pair. 

In the context of heavy-light systems with charm, the ALICE collaboration has published several results, including a study of the $\bar{D}N$ interaction~\cite{ALICE:2022enj}, and more recently, femtoscopy of $D^{(*)} \pi$, $D^{(*)} K$ and $D^{(*)} \bar{K}$ pairs~\cite{ALICE:2024bhk}. Theoretical predictions incorporating heavy-light pairs have been focused on the the neutral channels~\cite{Albaladejo:2023pzq,Liu:2023uly,Khemchandani:2023xup}, with the exception of~\cite{Torres-Rincon:2023qll} which presented predictions for the measured charged channels. While the sizable errors bars in the experimental data allow for a reasonable agreement across channels, the $D^+\pi^-$ case exhibits strong deviations between theory and experiment. While the experimental results suggest little to no interaction, the theoretical prediction indicates significant attraction, dominated by the lower pole of the $D_0^*(2300)$ state, which manifests as a shallow minimum in the correlation function. Incorporating the emission source for this broad resonance into the experimental data may improve the comparison. A proposed new channel where the molecular nature of the $D_0^*(2300)$ would more strongly influence the correlation function is the $D_s^+ K^-$ pair~\cite{Torres-Rincon:2023qll}.

\section{Studies at finite temperature}

Models that can dynamically generate states from an EFT description can be formally extended to finite temperature  using the imaginary-time formalism. This approach allows for the study of differences in the properties of molecular states compared to those in vacuum. The phenomenological relevance is motivated by measurements at RHICs, where a significant temperature is achieved.

\begin{figure*}
\centering
\vspace*{1cm}       
\includegraphics[width=8cm,clip]{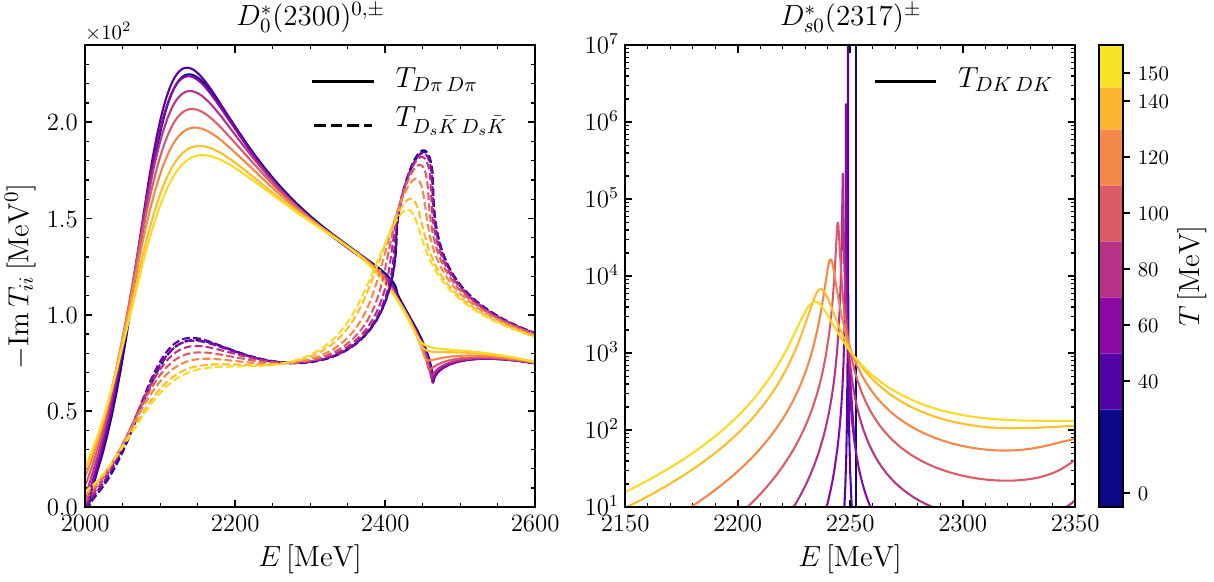}
\includegraphics[width=4cm,clip]{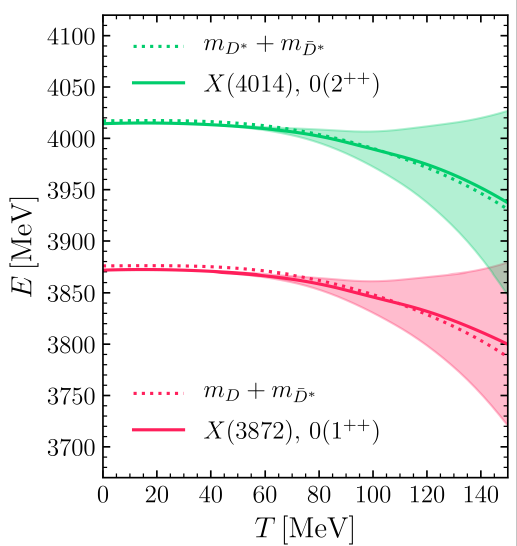}
\caption{Left panel: Temperature dependence of the spectral shapes of the two poles of the $D_0^*(2300)$ state. Middle panel: Same for the narrow $D_{s0}^* (2317)$ state. Right panel: Temperature dependence of the masses and decay widths of the $X(3872)$ and $X(4014)$. Figures taken from Refs.~\cite{Montana:2020lfi,Montana:2022inz}.}
\label{fig:thermal}  
\end{figure*}

In the charm sector, several works have examined---under different approximations---the thermal modification of the $D$-meson mass and decay width~\cite{Fuchs:2004fh,He:2011yi,Ghosh:2013xea,Cleven:2017fun,Torres-Rincon:2021yga}. A summary of these results can be found in the review~\cite{Das:2024vac}. A recent calculation has extracted the spectral function of $D$ and $D^*$ mesons using a self-consistent study~\cite{Montana:2020lfi,Montana:2020vjg}. As a byproduct, the properties of the generated molecular states $D_0^*(2300)$ and $D_{s0}^*(2317)$ can be monitored as a function of temperature. This is illustrated in the left and middle panels of Fig.~\ref{fig:thermal}, where the spectral shapes of the two poles of the $D_0^*(2300)$ and the $D_{s0}^*(2317)$ are shown. While all masses exhibit a small dependence on temperature up to $T=150$ MeV ($\Delta m \simeq -20$ MeV), the thermal widths acquire sizable values, particularly for the $D_{s0}^*(2317)$, which is bound in vacuum. This increase would alter their production in RHICs and enhance their decay rates.

In the hidden charm sector, molecular states such as $X(3872)$ and $X(4014)$ also change their nature from weakly bound states to thermal resonances as the temperature increases. As explained in~\cite{Montana:2022inz}, these states inherit the width of the $D$ and $\bar{D}^*$ mesons, of which they are composed. As shown in the right panel of Fig.~\ref{fig:thermal}, the dependence of their masses and widths varies only slightly with temperature between $T=0$ and $T \simeq 100$ MeV. However, for $T>150$ MeV, the mass of the exotics exceed the thermal masses of their constituents, and the decay widths (represented as a broad band around the mass) increase largely due to phase-space opening. Since these temperatures are relevant for RHICs, the resulting charm mesons from the decay of these exotics can be enhanced compared to vacuum expectations if the molecular hypothesis is correct.

\section{Conclusions}

Several features of some hadrons containing open and hidden charm, such as their proximity to two-particle thresholds and small decay widths, suggest their possible molecular nature. Effective field theory calculations demonstrate that this possibility is well founded through the explicit generation of these states with their correct properties (without excluding that other configurations, such as tetraquarks, may also play a role). Remarkable states, such as the $D_0^*(2300)$ and the narrow $D^*_{s0} (2317)$, can be described as heavy meson---light meson molecular states. Also, the exotic states $X(3872)$ and $X(4014)$ can be viewed as $D \bar{D}^*$ and $D^* \bar{D}^*$ bound states, respectively. 

Predictions at finite temperature---relevant for RHIC phenomenology---can also help to discriminate between different configurations. Explicit finite temperature calculations using EFTs show that the thermal bath modifies the properties of the molecular states---such as providing a finite decay to bound states, like the $D_{s0}^*(2317)$ and $X(3872)$---leading to changes in the predicted population of their decay products. This serves as an additional test for the molecular configuration hypothesis. Finally, charm femtoscopy has recently emerged as a valuable technique for studying the presence of these molecular states by analyzing specific characteristics of the correlation functions of their constituents.

\section{Acknowledgments}

This work has been supported by the project number
CEX2019-000918-M (Unidad de Excelencia ``Mar\'ia de Maeztu''), PID2020-118758GB-I00 and PID2023-147112NB-C21, financed by the Spanish MCIN/ AEI/10.13039/501100011033/, the
EU STRONG-2020 project, under the program H2020-
INFRAIA-2018-1 grant agreement no. 824093, and DFG project no. 315477589 - TRR 211 (Strong-interaction matter under extreme conditions).

\end{document}